\documentclass{article}
\usepackage{graphicx}
\usepackage{amsmath}
\usepackage{amsfonts}
\usepackage{amssymb}

\begin{document}

\begin{center}
{\Large Fast quantum algorithms for numerical integrals and stochastic
processes}\footnote{This work has been supported in part by a NDSEG
fellowship, by grant \# N00014-95-1-0975 from the Office of Naval Research, by
ARO and DARPA under grant \# DAAH04-96-1-0386 to QUIC, the Quantum Information
and Computation initiative, by a DARPA grant to NMRQC, the Nuclear Magnetic
Resonance Quantum Computing initiative, by the NASA/JPL Center for Integrated
Space Microsystems, and by the JPL\ Information and Computing Technologies
Research Section.}

\bigskip

Daniel S. Abrams and Colin P.\ Williams

\bigskip

Quantum Algorithms \& Technologies Group (QATG), Section 365

NASA Jet Propulsion Laboratory

California Institute of Technology

Pasadena, CA 91109-8099

\bigskip
\end{center}

We discuss quantum algorithms that calculate numerical integrals and
descriptive statistics of stochastic processes. With either of two distinct
approaches, one obtains an exponential speed increase in comparison to the
fastest known classical deterministic algorithms and a quadratic speed
increase in comparison to classical Monte Carlo (probabilistic) methods. We
derive a simpler and slightly faster version of Grover's mean algorithm,
demonstrate how to apply quantum counting to the problem, develop some
variations of these algorithms, and show how both (apparently quite different)
approaches can be understood from the same unified framework. Finally, we
discuss how the exponential speed increase appears to (but does not) violate
results obtained via the method of polynomials, from which it is known that a
bounded-error quantum algorithm for computing a total function can be only
polynomially more efficient than the fastest deterministic classical algorithm.

\newpage

\section{Introduction}

Quantum algorithms have been discovered that can solve many problems faster
than the best known classical algorithms. Most famous are Shor's factoring
algorithm\cite{Shor} and Grover's searching algorithm\cite{Grover}%
\cite{Grover2}, but quantum computers can also be used to simulate physics
with an exponential speedup\cite{AbramsEigen}, find means, and medians with a
quadratic speedup \cite{GroverFramework}, and solve a variety of artificial
problems \cite{DeutschJozsa}\cite{Simon} exponentially faster than is possible
classically. Still, due (perhaps) to the enormous technical challenges that
must be overcome before a useful quantum computer can ever be built, there is
a general sense that more applications must be found in order to justify
attempts to construct a quantum computing device.

We suggest one possible application of a quantum computer, namely, computing
the values of integrals. This problem can be solved in a fairly
straightforward manner via either quantum counting \cite{BrassardCounting}, or
Grover's mean estimation algorithm \cite{GroverFramework}. Although these
algorithms are not new, this application may be the most useful one described
to date. (Because N operations are required to retrieve N\ values from a
classical database, the mean finding algorithm affords no speed-up when
applied to a pre-existing data set. Indeed, even the original database search
algorithm has only limited utility, because it can only be used to search a
function space, not a true database. It not clear to how many real-life
problems it could be applied \cite{ZalkaRealDatabase}). We also show how these
apparently different algorithms can be understood from a unified perspective,
thereby explaining their equivalent computational complexity.

In addition, we suggest that a quantum computer may be used to determine
various characteristics of stochastic processes (for example, stock prices).
Frequently, such processes are used to generation distribution functions, and
one wishes to know the mean, variance, and higher moments. One can apply
quantum counting and mean estimation to obtain super-classical speedups for
these problems as well.

On a quantum computer, one can find the value of a $d$-dimensional integral in
O(1/$\epsilon$) operations, where $\epsilon$ is the desired accuracy. It
follows from the results of Nayak and Wu \cite{Nayak} that this is in fact a
lower bound. Classically, one requires O(1$/\epsilon^{2}$) operations to
achieve the same accuracy using probabilistic methods, and requires
O(1/$\epsilon^{d}$) \ - $exponentially$ more - operations to achieve the same
accuracy deterministically. (More precisely, it is polynomial in the accuracy
and exponential in the number of dimensions). Since real computers and all
classical devices are in fact deterministic, this exponential speed increase
is by no means a red herring. Indeed, there is a popular misconception that
real computers can perform probabilistic algorithms with impunity by employing
pseudo-random number generators. Of course, pseudo-random numbers are not
truly random at all - and one must in fact be careful about treating them as
such. For example, in 1992 Ferrenberg et al. found bugs in a supposedly good
pseudo-random number generator when a numerical simulation of an Ising spin
system failed due to hidden correlations in the ``random'' numbers
\cite{FerrenbergBadRandom}. The moral here is that one cannot rely upon a
classical computing device to properly execute a probabilistic algorithm. In
some (more than merely technical) sense, the quantum algorithm for evaluating
integrals provides an exponential speed increase.

The rest of this paper is organized as follows:\ first, we formalize the
problems and discuss the relevant classical algorithms. We then review
Grover's search algorithm from the more general perspective of amplitude
amplification. We describe two different approaches to mean estimation, one
using amplitude amplification and the other quantum counting, and provide a
new, simpler, and slightly faster version of the former algorithm. We discuss
some variations of these algorithms and show how they are essentially the
same. Finally, we conclude with a discussion of the lower bounds set by the
method of polynomials, and explain why they must be applied with greater care
than one might first suppose.

\bigskip

\section{Statement of the problem and classical algorithms}

Without loss of generality, we consider integrals of a real-valued
d-dimensional function $g(x_{1},x_{2},...x_{d})$\ defined for $x_{i}$ in the
range $[0,1]$ and where $g(x_{1},x_{2},...x_{d})\in\lbrack0,1]$, for all
values of $x_{i}$. Thus we seek to calculate
\begin{equation}
I=\int_{0}^{1}\int_{0}^{1}...\int_{0}^{1}g(x_{1},x_{2},...x_{d})dx_{1}%
dx_{2}...dx_{d} \label{Integral1}%
\end{equation}
In the discussion that follows, we shall approximate $g$ with a real-valued
d-dimensional function $f(a_{1},a_{2},...a_{d})$\ defined over integral values
$a_{i}$ in the range $[1,M]$ and where
\begin{equation}
f(a_{1},a_{2},...a_{d})=g(\frac{a_{1}}{M},\frac{a_{2}}{M},...\frac{a_{d}}{M})
\end{equation}
Thus, we wish to find the sum
\begin{equation}
S=\frac{1}{M^{d}}\sum_{x_{1}=1}^{M}\sum_{x_{2}=1}^{M}...\sum_{x_{d}=1}%
^{M}f(a_{1},a_{2},...a_{d}) \label{Sum1}%
\end{equation}

Note that the sum $S$ is identical to the average of $\ f$ over all $a_{i}$.
The accuracy with which the sum $S$ approaches the integral $I$\ is obviously
determined by the density of points $M$ in each variable and the shape of the
particular function. However\footnote{Because the computational complexity of
the quantum algorithms (and also the classical Monte Carlo algorithms, for
that matter) depend only logarithmically on $M,$ this approximation is not a
limiting factor (as long as the function is not pathological).}, in what
follows, our sole concern will be with approximating the sum $S$.

A sum of this form can also be used to determine properties of a stochastic
process. We describe a stochastic process by a sequence of values,
$w_{1},w_{2},...,w_{N}$, where each value $w_{i}$ is chosen randomly from a
distribution which may depend on some (or all) $w_{j}$ for $j<i$. For example,
a simple one dimensional random walk might be described by a sequence for
which each $w_{i}$ is either $(w_{i-1}+1)$ or $(w_{i-1}-1)$ with equal
probability. Often, one is interested in a property of such a sequence that
can be represented as a function $v(w_{1},w_{2},...,w_{N}).$\ (In many cases,
the function $v$ may depend only upon the final value $w_{N}).$ One wishes to
determine the mean, variance, skewness, and possibly higher moments of the
function $v$ over the space of all possible sequences. This problem is easily
transformed into the form (\ref{Integral1}) through a change of
variables:\ write each $w_{i}$ as a function $w_{i}(r_{i},w_{1},w_{2}%
,...,w_{i-1}),$ where $r_{i}$ is a random variable in the range [0,1]. Then we
can write $v$ as a function $v(r_{1},r_{2},...,r_{N})$ of the independent
random variables $r_{i\text{ }}$, scale the output so that it fits within the
desired range, and we have a function in the form $g$ above. The mean value of
the stochastic process is then simply the integral (\ref{Integral1}). Once
again, we represent the integral as a discrete sum. (For some stochastic
processes, the problem may in fact be discrete from the beginning). Thus the
problem again reduces to finding the sum $S$ in (\ref{Sum1}).

One can find higher moments of a stochastic process by simply applying the
above approach to a calculation of the mean of $v^{2},$ $v^{3},$ etc. This
method can of course also be applied to calculate moments of any distribution
function (even if it is not the result of a stochastic process) as long as it
can be represented in closed form.

It should be intuitively obvious that without any knowledge of the function
$f$, one requires classically $O(M^{d})$ operations to evaluate the sum. More
precisely, if we view $f$ as an oracle (or ``black-box''), then one requires
at least $M^{d}/2$ queries to determine $S$ to within $\pm\frac{1}{4}$.
\ (This is because it is possible that the remaining $M^{d}/2$ unqueried
function values may be either all 0's or all 1's, \ one of which will always
shift the mean by at least $\frac{1}{4}).$ It follows that an ordinary
classical Turing machine requires exponentially many operations to determine
$S$ with accuracy $\epsilon$\ for any $\epsilon<\frac{1}{4}.$

However, if one is allowed to employ a probabilistic algorithm, then one can
randomly sample values of the function $f$ for various $a_{1},a_{2},...a_{d}$;
as long as the values of $a_{i}$ are chosen randomly (and provided that you
are not exceedingly unlucky), it is possible to quickly approximate $S$ to any
desired precision. Indeed, it is a straightforward consequence of the central
limit theorem that one can determine $S$ with accuracy $\epsilon$ (with
bounded probability) using only $O(1/$ $\epsilon^{2})\,$\ operations. \ Note
that the number of trials does not depend at all upon the size of the
function's domain - as it did in the deterministic case - but only on the
desired accuracy. This is in fact how Monte Carlo integrals are computed, and
is essentially the only practical way to calculate integrals of functions with
high dimensionality. (It is also why we are not concerned with the
approximation of the integral $I$ with the sum $S$ - one can make $M $
essentially as large as one desires, paying only a logarithmic cost in
computational complexity). Unfortunately, Monte Carlo integrals on classical
devices require the use of a pseudo-random number generator, and as mentioned
previously, there is no guarantee that one will obtain ``good'' random
numbers. One obvious way to solve this dilemma would be to use a simple
quantum event to produce a string of truly random numbers; but once one
introduces quantum mechanics into the problem, we can find an even more
effective solution.

\section{Principle of Amplitude Amplification}

Both of the quantum algorithms discussed in this paper require a generalized
version of Grover searching. The treatment below follows that of Grover
\cite{GroverFramework}; similar ideas have also been described by Brassard et.
al. \cite{BrassardCounting} and various others.

All quantum algorithms consist of unitary operations applied in series. Any
sequence of unitary operations can be viewed as a single unitary operator.
Consider a particular unitary operator $U$ which has amplitude $U_{ts}$
\ between a starting state $|s\rangle$ and a target state $|t\rangle.$ If the
computer is initially in the state $|s\rangle,$ then after one application of
$U$ the computer will be found in the state $|t\rangle$ with amplitude
$U_{ts}$, and if the state of the computer is measured in the canonical basis,
the probability of obtaining the state $|t\rangle$ will therefore be $\left|
U_{ts}\right|  ^{2}$. We seek to $amplify$ the amplitude of the state
$|t\rangle.$ (Increasing the amplitude of this state increases the chances
that it will be found upon measurement and thereby allows for fast searching).

Amplitude amplification in its simplest form requires the inversion operator
$I_{x\text{ \ }}$which inverts the phase of the state $|x\rangle.$ We compose
the unitary operators $I$ and $U$ to form the unitary operator $G$ in the
following way:$.$%
\begin{equation}
G=-I_{s}U^{-1}I_{t}U
\end{equation}
It can be easily verified that the operator $G$ leaves invariant the subspace
spanned by $|s\rangle$ and $U^{-1}|t\rangle$. In particular, one finds that
\begin{equation}
G\left(  \alpha|s\rangle+\beta U^{-1}|t\rangle\right)  =\left\{  (1-4\left|
U_{ts}\right|  ^{2})\alpha+2U_{ts}\beta\right\}  |s\rangle+\left\{
-2U_{ts}^{\ast}\alpha+\beta\right\}  U^{-1}|t\rangle
\end{equation}
which is approximately a rotation by $2\left|  U_{ts}\right|  $ radians. It
follows that by applying $O(1/\left|  U_{ts}\right|  )$ iterations, one can
obtain the state $U^{-1}|t\rangle$ with near certainty.

The original fast searching algorithm \cite{Grover2} applies the above steps
with $U=W$, where $W$ is the Walsh-Hadamard transform - that is, a \ $\pi/2$
rotation of each qubit. If the initial state $|s\rangle$ = $|00...0\rangle, $
then $\left|  U_{ts}\right|  =\left|  W_{ts}\right|  =1/\sqrt{N}$ for all
possible target states $|t\rangle$. \ The unitary operation $I_{t}$
selectively inverts the phase of the actual target state $|t\rangle$ for which
we are searching. After one application of $W,$ the probability of measuring
$|t\rangle$ would be only $\frac{1}{N}$, the same as one would obtain
classically by guessing. However, it follows immediately from the above that
the amplitude $|t\rangle$ can be amplified to nearly 1 by applying only
$O(\sqrt{N})$ operations.

\bigskip

\section{Integrals via Amplitude Amplification}

To evaluate the integral $I$ in (\ref{Integral1}) (or alternatively the sum
$S$ in (\ref{Sum1})), one can use the mean estimation algorithm described by
Grover in \cite{GroverFramework}. We provide a simpler variation on this
algorithm, which distills the essential features from the original algorithm,
but eliminates unnecessary unitary operations and qubits while retaining the
essentials of the approach.

The algorithm works by refining a series of approximations. One can obtain an
intuitive understanding of the approach by employing an analogy to classical
coin-flipping:\ although it is tricky to describe, the algorithm is actually
quite simple in practice. Consider a coin, which, when tossed, comes up heads
with probability $p_{1}=S$. \ By the central limit theorem, one can determine
$S$ with accuracy $\delta$ using $O(1/$ $\delta^{2})\,$\ trials. Let us call
the first estimate so obtained $E_{1}$ and the error $\delta_{1}$. Then with
high probability $E_{1}-\delta/2<p_{1}<E_{1}+\delta/2.$ We now wish to ``zoom
in'' on this interval and determine $S$ more accurately within these bounds.
We thus define the difference $D$ $_{1}=S-(E_{1}-\delta/2),$ that is, the
distance that $S$ is from the bottom of the interval. Thus with high
probability $D_{1}$ is bounded by $0$ and $\delta$. We ``zoom in'' by
rescaling this value so that it is bounded by $[0,1]$ and call this value
$p_{2}$, that is $p_{2}=D_{1}/\delta.$

We now imagine a second iteration, where we are provided with a second coin
which lands heads with probability $p_{2}.$ Of course, one could not actually
make such a coin without knowing $S$ ahead of time, which would defeat the
purpose. However, in the analogous quantum problem it will be possible, so we
shall imagine that someone can in fact provide for us the coin with
probability $p_{2}$. $\ $As with $p_{1}$, one can determine $p_{2}$ with
accuracy $\delta$ using $O(1/$ $\delta^{2})\,$\ trials. Call this estimate
$E_{2}$; then $E_{2}-\delta/2<p_{2}<E_{2}+\delta/2$ with high probability.
However, because $p_{2}=D_{1}/\delta,$ an estimate of $p_{2}$ with accuracy
$\delta$ is an estimate of $D_{1}$ with accuracy $\delta^{2}.$ Since $S=D$
$_{1}+(E_{1}-\delta/2)$, we thus obtain an estimate of $S$ to accuracy
$\delta^{2}.$

We continue this process with further iterations. Let $D_{2}=S-(E_{1}%
-\delta/2)-(E_{2}-\delta/2)\delta,$ which is evidently bounded by
$[0,\delta^{2}];$ we define $p_{3}$ by rescaling $D_{2}$: that is, $p_{3}$ =
$D_{2}/\delta^{2}$, which is bounded by $[0,1\dot{]}$. We imagine a third coin
with probability of heads $p_{3}$ and determine this probability to accuracy
$\delta$ using $O(1/$ $\delta^{2})\,$\ trials, as before. But since $p_{3}$ =
$D_{2}/\delta^{2}$, this implies an estimate of $D_{2}$ to accuracy
$\delta^{3}$, which implies an estimate of $S$ to accuracy $\delta^{3}.$ We
then let $D_{3}=S-(E_{1}-\delta/2)-(E_{2}-\delta/2)\delta-(E_{3}%
-\delta/2)\delta^{2}$, etc. Each iteration requires the same number of coin
tosses, but improves our estimate by a factor of $\delta.$ With $O(n/$
$\delta^{2})$ tosses, we estimate $S$ with accuracy $\delta^{n}.$ Phrased
differently, the required number of coin tosses scales only as the log of the
desired accuracy. However, let us reiterate that this classical algorithm
could not actually be used in practice; it is discussed only to serve as an
analogy to the quantum algorithm, described below. It highlights the fact that
the final complexity of the algorithm will not be limited by the number of
trials (which we have seen is only logarithmic in the desired accuracy), but
by the fact that $O(1/$ $\epsilon)$ quantum logic operations are required to
prepare the final coin-like-state (that reveals $S$ to accuracy $\epsilon$).

We will now describe the quantum algorithm. Let $E$ be the current best
estimate for $S$; that is
\begin{equation}
E=(E_{1}-\delta/2)+(E_{2}-\delta/2)\delta+(E_{3}-\delta/2)\delta
^{2}+...+(E_{k}-\delta/2)\delta^{k-1}%
\end{equation}
As before, let $D=S-E$. We wish to obtain an estimate for $D$ with accuracy
$\delta^{k}$. To begin, we define a new function $f^{\prime}=f-E$. Recall
that
\begin{align}
D  &  =S-E\\
&  =\left(  \frac{1}{M^{d}}\sum\limits_{a_{1},a_{2},...a_{d}=0}^{M-1}%
f(a_{1},a_{2},...a_{d})\right)  -E\\
&  =\frac{1}{M^{d}}\sum\limits_{a_{1},a_{2},...a_{d}=0}^{M-1}f^{\prime}%
(a_{1},a_{2},...a_{d})
\end{align}
In other words, $D$ is the average value of $f$'. The essential quantum part
of the algorithm is to estimate the average value of $f$'; from this, we will
simply iterate to obtain finer estimates. To calculate the average, consider a
quantum computer with $d\log_{2}M+1$ qubits. Label the states $|r\rangle
|a_{1},a_{2},...a_{d}\rangle$ where the first qubit $r$ is a work qubit and
the remaining qubits indicate a value in the domain of $f$. The computer is
placed initially in the zero state: $|0\rangle|00....0\rangle.$ We begin by
applying a Walsh-Hadamard transform to the function qubits in order to obtain
an equal superposition of all possible values for the $a_{i}$:
\begin{equation}
\left|  \Psi_{1}\right\rangle =\frac{1}{\sqrt{M^{d}}}\sum\limits_{a_{1}%
,a_{2},...a_{d}=0}^{M-1}|0\rangle|a_{1},a_{2},...a_{d}\rangle
\end{equation}

Next we rotate the first qubit by an amount $f^{\prime}$. The state is then
\begin{equation}
\left|  \Psi_{2}\right\rangle =\frac{1}{\sqrt{M^{d}}}\sum\limits_{a_{1}%
,a_{2},...a_{d}=0}^{M-1}\sqrt{1-f^{\prime}(a_{1},a_{2},...a_{d})^{2}}%
|0\rangle|a_{1},a_{2},...a_{d}\rangle+f^{\prime}(a_{1},a_{2},...a_{d}%
)|1\rangle|a_{1},a_{2},...a_{d}\rangle
\end{equation}

Finally, we perform the inverse of the Walsh-Hadamard transform used in the
first step. It is easy to see that the amplitude of the state $|1\rangle
|00...0\rangle$ will then be $D$ (because each state $|1\rangle|a_{1}%
,a_{2},...a_{d}\rangle$ contributes amplitude $\frac{1}{\sqrt{M^{d}}}%
f^{\prime}(a_{1},a_{2},...a_{d})$ to the state $|1\rangle|00...0\rangle$). An
estimate for $D$ can therefore be obtained by making measurements of the state
of the system in repeated trials, and counting the frequency of the result
$|1\rangle|00...0\rangle.$ To obtain an accuracy $\epsilon$ requires $O(1/$
$\epsilon^{2})$ measurements.

However, we can use amplitude amplification to increase the accuracy of our
estimate. The steps described above can be viewed as a single unitary
operation $U$ that has amplitude $\left|  U_{ts}\right|  $ between the
starting state $|s\rangle=|0\rangle|00...0\rangle$ and the target state
$|t\rangle=|1\rangle|00...0\rangle$. It follows that one can use amplitude
amplification to increase the probability of measuring the state
$|1\rangle|00...0\rangle$. By performing only $O(N)$ operations, one can
increase the amplitude of $|t\rangle$ to $N\ast D$.\ The largest value one can
choose for $N$ is $O(1/\delta^{k})$ (because the magnitude of $D$ is
$O(\delta^{k})$ and the norm of the amplified amplitude is bounded by one). In
this case, the probability amplitude for the target state is then roughly
$D/\delta^{k}$\footnote{Because the amplitude amplification process is not
exactly linear, the final amplitude is not exactly\ $D/\delta^{k}$.\ However,
this difference can be easily accounted for and has no effect on the
computational complexity of the algorithm.}. Borrowing from our earlier
classical analogy, this is the scaled (``zoomed-in'') value $p_{k+1}$. With
the same $O(1/$ $\epsilon^{2})$ trials, we thus determine $p_{k+1}$with
accuracy $\epsilon$, but this provides an estimate of $D$ with accuracy
$\epsilon^{\ast}=$ $\epsilon/N=\epsilon\delta^{k}$. \ If we vary $N$ and fix
$\epsilon$, we perform only $O(1/\epsilon^{\ast})$ operations to find $D$ with
accuracy $\epsilon^{\ast}.$

Because of the limit on the size $N$, the algorithm requires several
iterations. Initially, $\ D$ may be any value between 0 and 1, and hence $N$
can be at most 1.\ (That is, we cannot use amplitude amplification at all). As
the estimates become more accurate, then the value of $D$ becomes
correspondingly smaller, and one can choose larger and larger $N.$ \ 

Just as in the classical case, each estimate $p_{k}$ \ is determined with a
fixed number of trials, but since the corresponding estimates of $D$ (and $S)$
become exponentially more accurate with each iteration, the total number of
trials is only a logarithmic function of the desired accuracy. Hence, the
significant contribution to the computational complexity is not the number of
trials. Instead, the complexity is determined by the amplitude amplification.
Within a polylogarithmic factor, the entire cost of the algorithm occurs on
the last iteration (because each iteration takes exponentially more time). The
computational complexity of the entire algorithm is therefore the same as the
amplitude amplification of the last iteration:\ $O(1/$ $\epsilon^{\ast})$
operations are required, where $\epsilon^{\ast}$ is the desired accuracy.

It is interesting to note that, as with the classical Monte Carlo method, the
quantum algorithm depends only upon the desired accuracy:\ the size of the
function's domain ($M^{d}$) is irrelevant.

\section{Integrals via Quantum Counting}

There is another algorithm which can be used to evaluate the sum $S$ in
(\ref{Sum1}), based upon the idea of quantum counting \cite{BrassardCounting}.
To use this method, one must first convert the real-valued function
$f(a_{1},a_{2},...a_{d})$ into a boolean valued function. This is accomplished
via the addition of an extra parameter $q$. The parameter takes on integral
values in the range [1,$Q$] where $Q$ is determined by the desired accuracy.
We then define
\begin{equation}
b(a_{1},a_{2},...a_{d},q)=
\begin{array}
[c]{c}%
1\text{ if }q\leq f(a_{1},a_{2},...a_{d})\ast Q\\
0\text{ if }q>f(a_{1},a_{2},...a_{d})\ast Q
\end{array}
\end{equation}
In other words, for a given $a_{1},a_{2},...a_{d},$the fraction of the $Q$
values for which $b(a_{1},a_{2},...a_{d},q)=1$ is the best approximation to
$f(a_{1},a_{2},...a_{d}).$ It follows that the average value of $b\,\ $is
identical to the average value of $f$. However, since $b$ is a boolean-valued
function, one can estimate the average value of $b$ via approximate counting.
That is, $S=\left\langle b\right\rangle =\frac{r}{M^{d}Q}$, where $r$ is the
number of solutions $b(a_{1},a_{2},...a_{d},q)=1$. To count the number of
solutions $r$, we recall that during the amplitude amplification process, the
state of the system rotates within the subspace spanned by $|s\rangle$ and
$U^{-1}|t\rangle$ at a rate which is proportional to $\left|  U_{ts}\right|
$. Moreover, we recall that by using the Walsh-Hadamard transform for $U$ (as
in the Grover search algorithm), the magnitude of $U_{ts}$ is exactly $\left|
U_{ts}\right|  =\left|  W_{ts}\right|  =1/\sqrt{N}$ for any given target state
$|i\rangle$. But if the target state is a sum over all basis vectors for which
$b(i)=1$, that is, $|t\rangle=\sum\limits_{i\in b(i)=1}|i\rangle$, then the
amplitude of $\left|  U_{ts}\right|  =\sqrt{r/N}$. Hence the amplitudes of the
states $|s\rangle$ and $U^{-1}|t\rangle$ will oscillate with a frequency that
depends on $r$. It is therefore a simple matter to create a superposition
\begin{equation}
|\Psi>=\frac{1}{\sqrt{A}}\sum\limits_{j=0}^{A-1}|j\rangle G^{j}|s\rangle
\end{equation}
and determine the value of $r$ by performing a fast Fourier transform on the
first register. The accuracy $1/A$ will depend linearly upon the number of
points used in the FFT, as will the number of quantum logic operations
(because it takes $O(1)$ operations to perform $G$, one requires $O(A)$
operations to create the state $|\Psi\rangle$ above). It follows that one can
determine the value of the integral $\ f$ $\ $to accuracy $\epsilon$ with
$O(1/\epsilon)$ operations, as in the previous algorithm. Also as above, we
find that the number of operations does not depend upon the size of the domain
of $\ f$, but only upon the desired accuracy.

\section{Discussion}

At first, it may appear surprising that these two very different quantum
algorithms should both require $O(1/\epsilon)$ operations. However, by
exploring some variations of these algorithms, we find that, while not
identical, they are both quite similar.

First, we note that there is a trivial variation of quantum counting, which is
simply to measure the state of the system in repeated trials, and count the
number of times one obtains the target state (or more precisely, a state for
which $b(a_{1},a_{2},...a_{d},q)=1$.) That is, we determine the fraction
$\frac{r}{M^{d}Q}=\left\langle b\right\rangle =S$ through random sampling.
This technique is directly analogous to the way, in Grover mean estimation, we
find the probability $p_{1\text{ }}$through repeated trials (counting the
number of times we measure the target state $|1\rangle|00...0\rangle$). In
both cases, $O(1/\epsilon^{2})$ operations would be required to obtain an
accuracy $\epsilon$. The difference is that using the Grover method, one can
subtract the most recent estimate from each term in the sum (to obtain the
function $f^{\prime}$), and then perform amplitude amplification to increase
the probability of obtaining the target state. By amplifying this difference,
the precision of the algorithm is limited by the (nearly)\ linear amplitude
amplification process rather than by the quadratic sampling process. In the
case of quantum counting, one can also apply the amplitude amplification
process to the target state (indeed, this is exactly what the quantum counting
algorithm does). However, one cannot subtract the most recent estimate from
each term in the sum:\ specifically, for a given $a_{1},a_{2},...a_{d}$, there
can be no less than zero values of $q$ for which $b(a_{1},a_{2},...a_{d}%
,q)=1.$\ \ In the Grover method, individual terms in the sum may be negative,
even though the sum of all the terms is always positive. The counting method
does not allow this possibility. It is therefore impossible to use the
technique of iterated, refined estimates to increase the precision of the approximation.

The relationship can be viewed from another perspective by considering a
variation of Grover's method. As presented earlier, the technique depends upon
measuring the amplitude of the target state $|1\rangle|00...0\rangle$. This is
accomplished through repeated measurements. However, one can also determine
this amplitude with a quantum FFT. Recalling once again that during the
amplitude amplification process the state of the system rotates within the
subspace spanned by $|s\rangle$ and $U^{-1}|t\rangle$, at a rate which is
proportional to $\left|  U_{ts}\right|  $ (which in this case is equal to
$p_{k}$), we see that one could also use an FFT to determine $\left|
U_{ts}\right|  $ (and therefore $p_{k}$). As in the case of quantum counting,
one requires $O(1/\epsilon)$ operations to obtain the result with accuracy
$\epsilon$. Moreover, because the FFT measures the frequency of the rotation,
one does not need to perform the iterated estimates (which previously ensured
that the initial amplitude $\left|  U_{ts}\right|  $ was sufficiently small
that it would in fact be amplified throughout the entire process).

The situation is in many ways similar to the relationship between Shor's
algorithm and Kitaev's algorithm\cite{Kitaev}. In the Kitaev algorithm, one
estimates the phase of an eigenvalue $\phi$ of a unitary operator $\widehat
{U}$ . The number of operations required to estimate $\phi$ grows polynomially
with the desired precision, but Kitaev obtains exponential precision by
considering $\widehat{U}^{2},\widehat{U}^{4},\widehat{U}^{8},$etc. This
process is analogous to the refined estimates used in the Grover method. In
\cite{Cleve}, Cleve et. al. describe how to modify Kitaev's algorithm so that
it uses an FFT to estimate the phase. The resulting algorithm is then
identical to Shor's.

A final variation of these mean finding algorithms arises naturally from the
following considerations\footnote{This last variation on quantum counting was
suggested by Peter Hoyer.}. In (our version of)\ the Grover algorithm, we
apply the unitary operators $W^{-1}RW$, where $W$ is the Walsh-Hadamard
transformation and $R$ is the rotation by $f^{\prime}$ (which maps $\left|
\Psi_{1}\right\rangle $ to $\left|  \Psi_{2}\right\rangle $ in the previous
description). The three unitary operators $W^{-1}RW$ take the initial zero
state into a target state with an amplitude proportional to $D$, the number we
seek to estimate. In the quantum counting algorithm, we begin with the zero
state, and apply only the operator $W$ to obtain a (different) target state,
also with an amplitude that is proportional to the (square root of
the)\ number we wish to estimate. In the final variation, we apply the
operators $\widehat{R}W$: \ that is, we leave out the final inverse
Walsh-Hadamard, and (to compensate)\ use a slightly different rotation
$\widehat{R}.$ Because it is the rotation $R$ that shifts the amplitude
according to the function we are trying to estimate, the final $W$ is in some
sense an extra, unnecessary step. However, if we use the original rotation $R$
from the modified Grover algorithm and consider the target state to be all
those states where the first qubit is $|1\rangle$, the amplitude of the target
state would then be proportional to the sum (or mean) of all values of $f$
squared -- which is not quite what we desire. Hence we simply perform a
modified rotation $\widehat{R}.$ which rotates by $\sqrt{f}$, in which case
the target state will occur with probability proportional to the sum\ of all
values of $\sqrt{f}$ squared, i.e., to the mean of $f$. \ By repeated
iterations of this process, we can perform an FFT (as in quantum counting) and
determine the mean of $f$ with the same linear scaling that we obtain with the
other approaches.

It is worth noticing that in this last variation, as with the original quantum
counting, one cannot use the method of iterated estimates (like we did in the
original Grover technique)\ because -- once again -- it is not possible to
account for negative values of $f^{\prime}$. It is also interesting to compare
this method with the original algorithms and ask why we need to introduce the
square root?\ With respect to the original Grover technique, this is because
of the difference between summing and then squaring (as we do in the original
algorithm) and squaring and then summing (which we do in the later algorithm).
However, in the case of quantum counting, the amplitude of the final state is
determined by squaring and then summing as well. But since the original
quantum counting applies to boolean values, all components of the
superposition occur with equal weight, and the result is the same.

The algorithms described above are summarized by the following chart:%

\begin{equation}%
\begin{array}
[c]{llllll}%
\mathbf{|s\rangle} & \mathbf{|t\rangle} & \mathbf{U} & \left|  U_{ts}\right|
& \mathbf{Method} & \mathbf{Complexity}\\
|0\rangle|00....0\rangle &
\begin{array}
[c]{l}%
\\
|1\rangle|00....0\rangle\\
\text{ \ }%
\end{array}
& W^{-1}RW & S &
\begin{array}
[c]{l}%
\text{Sampling with }\\
\text{iterated estimates}%
\end{array}
& O(1/\epsilon)\\
|0\rangle|00....0\rangle &
\begin{array}
[c]{l}%
\\
|1\rangle|00....0\rangle\\
\text{ \ }%
\end{array}
& W^{-1}RW & S & \text{FFT} & O(1/\epsilon)\\
|00....0,0\rangle &
\begin{array}
[c]{l}%
\\
\sum\limits_{i\in b(i)=1}|i\rangle\\
\text{ \ }%
\end{array}
& W & \sqrt{S} &
\begin{array}
[c]{l}%
\text{Sampling }\\
\text{(no iterated estimates)}%
\end{array}
& O(1/\epsilon^{2})\\
|00....0,0\rangle &
\begin{array}
[c]{l}%
\\
\sum\limits_{i\in b(i)=1}|i\rangle\\
\text{ \ }%
\end{array}
& W & \sqrt{S} & \text{FFT} & O(1/\epsilon)\\
|0\rangle|00....0\rangle &
\begin{array}
[c]{l}%
\\
\sum\limits_{a_{i}=0}^{M-1}\sqrt{f}|1\rangle|a_{1},a_{2},...a_{d}\rangle\\
\text{ \ }%
\end{array}
& \widehat{R}W & \sqrt{S} &
\begin{array}
[c]{l}%
\text{Sampling }\\
\text{(no iterated estimates)}%
\end{array}
& O(1/\epsilon^{2})\\
|0\rangle|00....0\rangle &
\begin{array}
[c]{l}%
\\
\sum\limits_{a_{i}=0}^{M-1}\sqrt{f}|1\rangle|a_{1},a_{2},...a_{d}\rangle\\
\text{ \ }%
\end{array}
& \widehat{R}W & \sqrt{S} & \text{FFT} & O(1/\epsilon)
\end{array}
\end{equation}

We see, therefore that the two apparently distinct algorithms are in fact both
very closely related. In both cases, we perform a sequence of unitary
operations that generate an operator with amplitude $\left|  U_{ts}\right|  $
to make a transition from the $\left|  0\right\rangle $ state to the target
state $\left|  t\right\rangle $, where the value of $\left|  U_{ts}\right|  $
depends directly on the sum $S$. \ In both cases, we can use a quantum FFT to
estimate the value of $\left|  U_{ts}\right|  $ and approximate $S$ with
accuracy $\epsilon$ in $O(1/$ $\epsilon)$ operations. In both cases, we can
estimate the value of $\left|  U_{ts}\right|  $ directly through repeated
measurements and then approximate $S$ with accuracy $\epsilon$ in $O(1/$
$\epsilon^{2})$ operations. The only difference is that in Grover's method,
the particular form of the operator $U$ allows one to consider negative values
$f^{\prime}$, which in turn allows one to use the process of iterated, refined
estimates and thus to obtain linear precision directly with repeated
measurements instead of with the fast Fourier transform.

\section{ Conclusion}

To briefly summarize:\ we have proposed two new applications for quantum
computation:\ evaluating integrals and calculating descriptive statistics of
stochastic process. Whereas $O(M^{d})$ operations are required on a classical
deterministic Turing machine, and $O(1/$ $\epsilon^{2})$ operations are
required with a classical probabilistic algorithm, one can obtain the same
accuracy on a quantum computer with only $O(1/$ $\epsilon)$ quantum
operations, using two different algorithms. We have provided a simpler (and
slightly more efficient) version of Grover's mean-finding algorithm,
demonstrated how quantum counting can be applied to mean estimation, derived
some variations of both algorithms, and shown how the two are very closely related.

In concluding, we would like to make two points. The first is that, while
these algorithms are probabilistic in nature, the mean estimation algorithms
employing FFTs do not rely upon sampling the function space, as do classical
Monte Carlo methods. The quantum algorithms in some sense consider the entire
(exponentially large) domain of the function all in one shot, and, with high
probability, return the mean to within the desired accuracy.

Second, it is interesting to consider our results in light of the work by
Beals et. al. \cite{Beals}, where it is proven (using the method of
polynomials)\ that a bounded-error quantum algorithm for computing a total
function can be only polynomially more efficient than the fastest
deterministic classical algorithm. \ A boolean function $b(a_{1}%
,a_{2},...a_{d},q)$ such as the one described in Section 5 can be described as
a sequence of $M^{d}q$ boolean values; the average of $b$ is a function of
those $M^{d}q$ boolean values, and it is a total function, since it is
well-defined for all possible input functions $b$. In order to phrase
mean-estimation as a decision problem, we can ask:\ ``Is the average value of
$b$ within the range $[E-$ $\epsilon,E+$ $\epsilon]$ ?'' (for some chosen $E$
and $\epsilon$). Naively, it appears that the results of \cite{Beals} would
imply that this problem cannot be speed up more than polynomially on a quantum
computer (vs. a classical deterministic computer) - whereas we have just
finished demonstrating an exponential separation. It appears that there is a
contradiction.\footnote{Actually, this issue applies equally to the
exponential separation between the classical deterministic and probabilistic algorithms.}

The (in fact quite simple) resolution of this problem is that the decision
question posed above does not quite correspond to mean-estimation. According
to the question given, a function with mean just slightly (infinitesimally)
more than $E+$ $\epsilon$ does not have a mean that is approximately $E$,
whereas a function that has mean exactly $E+$ $\epsilon$ does. Of course, our
quantum algorithms cannot reliably differentiate between these two cases in
polynomial time any better than the classical deterministic algorithms can.
The decision question that one can associate with mean-estimation would be a
probabilistic one; the answer should be sometimes yes and sometimes no with a
probability that depends (perhaps as a gaussian function) upon the distance
the true mean is from the estimate $E$. \ Such a question is not a function
(although it can be viewed as the average value of a weighted ensemble of
functions). Thus, the results obtained in \cite{Beals} do not apply to our
problem, and there is no contradiction.

In concluding therefore the authors would like to make the following point. It
is easy for results such as those in \cite{Beals} to cause one to be
disheartened about the prospects of quantum computing. However, sometimes the
``real'' problems we wish to solve have special properties that can make them
easier than the general cases. Calculating approximate integrals is one such
example - and there are likely others waiting to be discovered.

\section{Acknowledgments}

D.S.A. gratefully acknowledges support from a NDSEG fellowship and from
NASA-JPL, and helpful conversations with Peter Hoyer and Seth Lloyd. Portions
of this research were supported by grant \# N00014-95-1-0975 from the Office
of Naval Research, by ARO and DARPA under grant \# DAAH04-96-1-0386 to QUIC,
the Quantum Information and Computation initiative, by a DARPA grant to NMRQC,
the Nuclear Magnetic Resonance Quantum Computing initiative, by the NASA/JPL
Center for Integrated Space Microsystems, and by the JPL\ Information and
Computing Technologies Research Section.


\newpage

\begin{thebibliography}{99}
\bibitem{AbramsEigen}D.S. Abrams and S. Lloyd, sub. to Phys. Rev. Lett., quant-ph/9807070

\bibitem {Beals}R. Beals, H. Buhrman, R. Cleve, M. Mosca, R. de Wolf, in
Proceedings of the 39th Annual Symposium on Foundations of Computer Science (1998)

\bibitem {BrassardCounting}G. Brassard, P. Hoyer, A. Tapp, quant-ph/9805082

\bibitem {Cleve}R. Cleve, A. Ekert, C. Macchiavello, M. Mosca, Proc. R. Soc.
Lond. A 454 (1998), pp. 339-354

\bibitem {DeutschJozsa}D. Deutsch and R. Jozsa, Proc. R. Soc. Lond. A 439, 553 (1992)

\bibitem {FerrenbergBadRandom}A. M. Ferrenberg and D. P. Landau,
Phys.\ Rev.\ Lett. 69, 23 (1992) pp3382-3384

\bibitem {Grover}L.K. Grover, Proceedings of the Twenty-Eighth Annual ACM
Symposium on the Theory of Computing, p. 661, 212-19

\bibitem {Grover2}L.K. Grover, Phys.\ Rev. Lett. 79, 325-328 (1997)

\bibitem {GroverFramework}L.K. Grover, quant-ph/9711043

\bibitem {Kitaev}A. Yu. Kitaev, quant-ph/9511026

\bibitem {Nayak}A. Nayak and F. Wu., quant-ph/9804066

\bibitem {Shor}P. Shor, in Proceedings of the 35th Annual Symposium on
Foundations of Computer Science, edited by S. Goldwasser (IEEE\ Computer
Society, Los Alamitos, CA, 1994), p.124

\bibitem {Simon}D. Simon, Proceedings of the 35th Annual Symposium on
Foundations of Computer\ Science, edited by S. Goldwasser (IEEE\ Computer
Society, Los Alamitos, CA, 1994), p.116

\bibitem {ZalkaRealDatabase}C. Zalka, quant-ph/9901068
\end{thebibliography}
\end{document}